\documentclass[epj]{webofc}
\usepackage[utf8]{inputenc}
\usepackage[varg]{txfonts}   
\usepackage{booktabs}
\usepackage{xcolor}
\definecolor{darkred}{rgb}{0.4,0.0,0.0}
\definecolor{darkgreen}{rgb}{0.0,0.4,0.0}
\definecolor{darkblue}{rgb}{0.0,0.0,0.4}
\usepackage[bookmarks,linktocpage,colorlinks,
    linkcolor = darkred,
    urlcolor  = darkblue,
    citecolor = darkgreen]{hyperref}
\wocname{EPJ Web of Conferences}
\woctitle{Lattice2017}
\begin{document}
%
\selectlanguage{english}
\title{%
Sanity check for $NN$ bound states in lattice QCD with L\"uscher's finite volume formula\\
-- Exposing  Symptoms of  Fake Plateaux --
}
\author{%
\firstname{Sinya} \lastname{Aoki}\inst{1,2}\fnsep\thanks{Speaker, \email{saoki@yukawa.kyoto-u.ac.jp} }\fnsep\thanks{This work is supported in part by the Grant-in-Aid of the Japanese Ministry of Education, Sciences and Technology, Sports and Culture (MEXT) for Scientific Research (Nos. JP15K17667, JP16H03978),  
by a priority issue (Elucidation of the fundamental laws and evolution of the universe) to be tackled by using Post ``K" Computer, 
and by Joint Institute for Computational Fundamental Science (JICFuS). We thank other members of HAL QCD collaboration for useful discussion.}
\and
\firstname{Takumi} \lastname{Doi}\inst{3,4} \and
\firstname{Takumi}  \lastname{Iritani}\inst{3}
}
\institute{%
Center for Gravitational Physics, Yukawa Institute for Theoretical Physics, Kyoto University,
Kitashirakawa Oiwakecho, Sakyo-ku, 
Kyoto 606-8502, Japan
\and
Center for Computational Sciences, University of Tsukuba, Tsukuba 305-8577, Japan
\and
 Theoretical Research Division, Nishina Center, RIKEN, Wako 351-0198, Japan
 \and
 iTHEMS Program, RIKEN, Wako 351-0198, Japan
}
\abstract{%
A sanity check rules out certain types of obviously false results, but does not catch every possible error. 
After reviewing such a  sanity check for $NN$ bound states with the L\"uscher's finite volume formula\cite{Iritani:2016jie,Aoki:2016dmo,Iritani:2017rlk},
we give further evidences for the operator dependence of plateaux, a symptom of the fake plateau problem, against the claim in \cite{Yamazaki:2017euu}. 
We then present our critical comments  on \cite{Beane:2017edf} by NPLQCD:
(i) Operator dependences of plateaux in NPL2013\cite{Beane:2012vq, Beane:2013br}
exist with the $P$-values of 4--5\%. 
(ii) The volume independence of plateaux in NPL2013 does not prove their correctness.
(iii) Effective range expansion (ERE) fits in NPL2013 violate the physical pole condition.
(iv) Ref.~\cite{Beane:2017edf} is partly based on new data and analysis different from the original ones\cite{Beane:2012vq, Beane:2013br}. (v) A new ERE in Refs.~\cite{Beane:2017edf,Wagman:2017tmp} does not satisfy the L\"uscher's finite volume formula.
}

\maketitle

\section{Introduction}\label{intro}
In the previous publications\cite{Iritani:2016jie,Aoki:2016dmo,Iritani:2017rlk}, we pointed out that the binding energy of the two nucleon ($NN$) system at heavy pion masses
extracted from the temporal correlation function at $t\simeq 1 $ fm  in lattice QCD is unreliable due to contaminations of excited scattering states.  
Tab.~\ref{tab:Summary} summarizes  our sanity check of $NN$ data in recent literatures, 
all of which employ the plateau fitting at $t\simeq 1 $ fm.
The sanity check is performed to rule out certain classes of obviously false results, but
the correctness is not guaranteed even if the data pass the test.  
The fact that none in the table passes the test
brings a serious doubt on existence of $NN$ bound states in $^1S_0$ or $^3S_1$ channels claimed on the basis of these data.
More sophisticated method than the simple plateau fitting, such as the variational method\cite{Luscher:1990ck}, is mandatory for the reliable study of  $NN$ systems.
\begin{table}[bth]
\vskip -0.2cm
\centering
\caption{A summary of our assessment (Source independence,
(i) consistency (ii) regularity (iii) physical pole).  Here ``No" means that the check fails while $\dagger$ implies no study on the item.  See Ref.~\cite{Iritani:2017rlk} for more details. }  
\label{tab:Summary}
\begin{tabular}{|c||c|c|c|c|c|c|c|c|} 
\hline
& \multicolumn{4}{c|}{$NN(^1S_0)$} & \multicolumn{4}{c|}{$NN(^3S_1)$} \\ 
\hline
 Data & Source & \multicolumn{3}{c|}{Sanity check} & Source & \multicolumn{3}{c|}{Sanity check} \\
     & independence & (i) & (ii) & (iii) & independence  & (i)  & (ii) & (iii) \\
\hline\hline
YKU2011 \cite{Yamazaki:2011nd}  & $\dagger$ &   No        &   No       &       &$\dagger$&  No          &    No     &     \\  
YIKU2012 \cite{Yamazaki:2012hi}  & No            &$\dagger$&  No       &        &  No         & $\dagger$ &   No     &     \\
YIKU2015 \cite{Yamazaki:2015asa}  & $\dagger$ & $\dagger$&  No      &        &$\dagger$&  $\dagger$ &   No     & No    \\
NPL2012 \cite{Beane:2011iw}  & $\dagger$ & $\dagger$ &   No     &        &   $\dagger$& $\dagger$ &             &          \\
NPL2013 \cite{Beane:2013br,Beane:2012vq}   & No            &                 &            &No    &   No            &        &             &  No         \\
NPL2015 \cite{Orginos:2015aya}   & $\dagger$ &   No         &            & No    &   $\dagger$& No   &            & No    \\
CalLat2017 \cite{Berkowitz:2015eaa} & No             &  ? &   & No     &  No             & ?   &             &  No \\
\hline
\end{tabular}
\vskip -0.4cm
\end{table}

Our concerns mentioned in Refs.~\cite{Iritani:2016jie,Aoki:2016dmo,Iritani:2017rlk}, however,  seem to be misunderstood by authors in Refs.~\cite{Yamazaki:2017euu,Beane:2017edf}.
Therefore we will first emphasize again the serious problem of the plateau fitting for the $NN$ system and then provide critical comments on Refs.~\cite{Yamazaki:2017euu,Beane:2017edf} in the remaining part of this report.

\section{Fake plateau problem}\label{sec-1}
\vskip -0.6cm
\begin{figure}[bth]
\centering
  \includegraphics[width=0.47\textwidth]{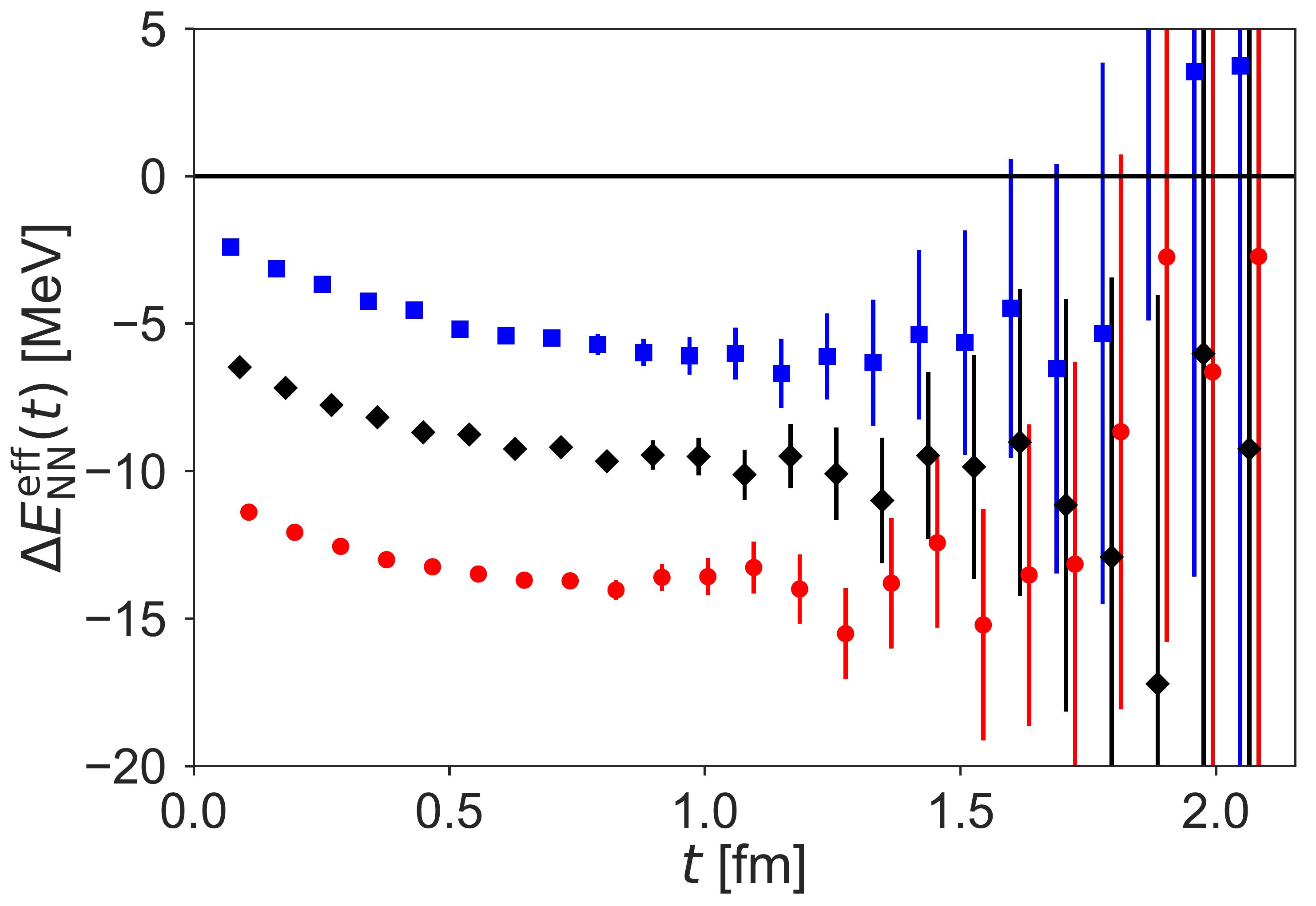}
  \includegraphics[width=0.43\textwidth]{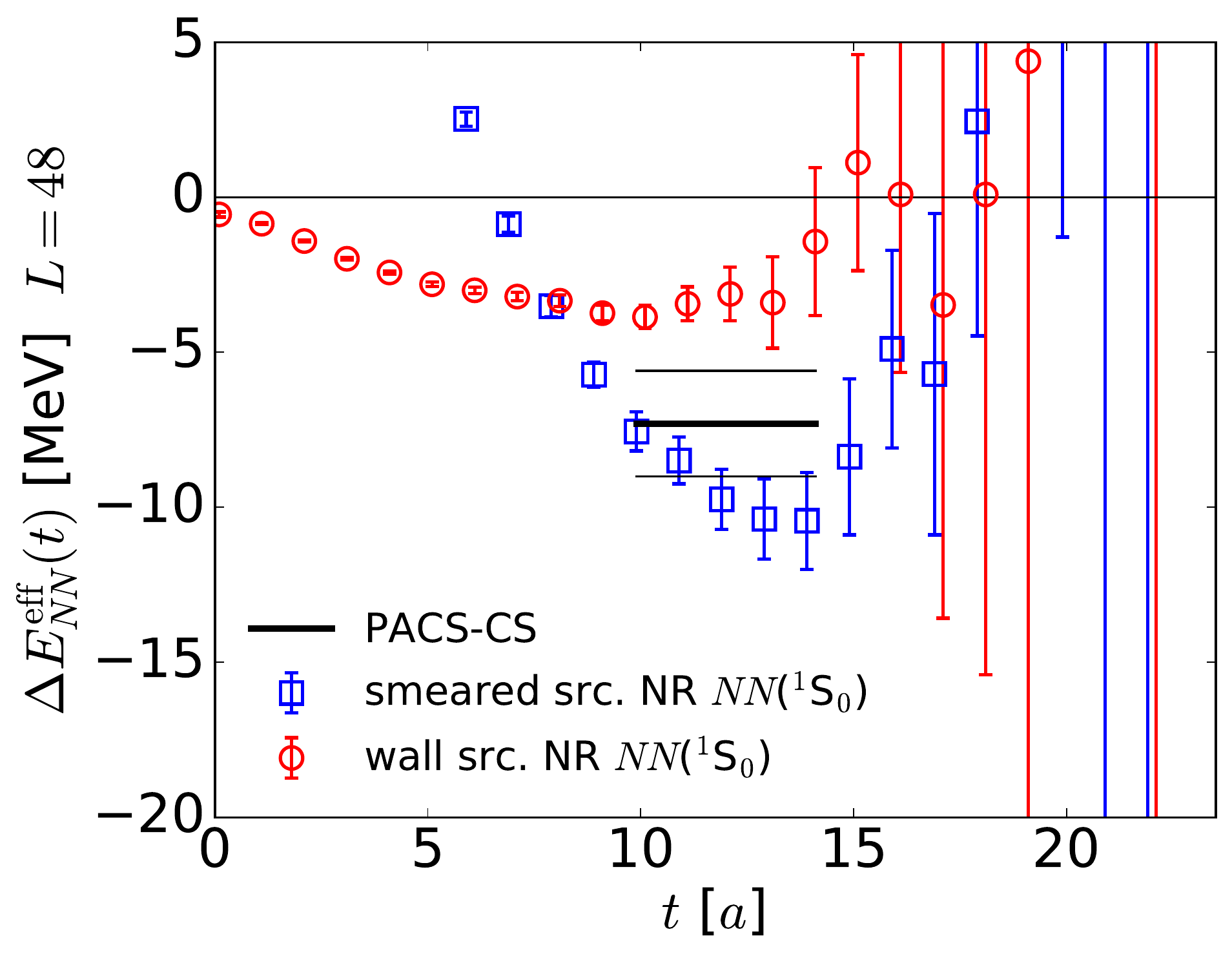}
 \caption{
(Left)  $\Delta E^{\rm eff}_{NN}(t) $ with fluctuations and errors  as a function of 
   $t$. 
(Right)   $\Delta E_{NN}^{\rm eff}(t)$ from the wall source (red circles) and the smeared source (blue squares)\cite{Iritani:2016jie}.
}
 \label{fig:demo}
 \vskip -0.5cm
\end{figure}
Plateaux of the $NN$ effective energy shift at early $t\simeq 1$ fm  cannot be trusted in general,
since the fake plateaux can easily appear due to contaminations from elastic excited states.
To demonstrate this, we consider the mock-up data for the effective energy shift\cite{Iritani:2016jie}, described by
\begin{eqnarray}
\Delta E^{\rm eff}_{NN}(t) = -\frac{1}{a}\log\left(\frac{R(t+a)}{R(t)}\right), \quad
R(t) = e^{-\Delta E_{NN} t} \left( 1 + b_1 e^{-\delta E_{\rm el.} t} + c_0  e^{-\delta E_{\rm inel.} t} \right).
\end{eqnarray}
with the correct energy shift $\Delta E_{NN}$,  
where we take $\delta E_{\rm el.} = 50$ MeV, 
which is the typical lowest elastic excitation energy at $L\simeq 4$ fm (and $m_N \simeq 2$ GeV), and $\delta E_{\rm inel.}= 500$ MeV, which is the mass of the heavy pion.
In this situation,
we naively expect  that the ground state saturation requires $t \ge 1/\delta E_{\rm el.}  \simeq 4$ fm,
which however is too large to have good signals for $NN$ systems.
The strategy used in the previous studies 
is tuning the source operators to satisfy $b_1\simeq 0$,
so as to achieve the ground state saturation at much smaller $t$ before noises dominate.
In practice, the source/sink operator is tuned  to have 
a plateau in $\Delta E^{\rm eff}_{NN}(t)$ at much smaller $t$.

Fig.~\ref{fig:demo} (Left) shows
three examples of $\Delta E^{\rm eff}_{NN}(t)$ with $b_1=0$ (an optimally tuned operator)
and $b_1=\pm 0.1$  keeping $c_0=0.01$,  where random fluctuations are assigned to $R(t)$ whose magnitude increases exponentially in $t$.
Indeed the fake plateaux appear at $t\simeq 1$ fm.
If no legend of data is given as in the figure,
we can not tell which are fake from the $t$ dependence. 

This example clearly demonstrates that the strategy mentioned above does NOT work at all:
Even if the plateau appears in $\Delta E^{\rm eff}_{NN}(t)$,
we cannot tell  whether it is fake or not.
Hereafter we will use the words,  ``the fake plateau problem", to remind readers of this fundamental problem.
This is the serious issue  for the current $NN$ data and nothing more is needed to doubt the validity of data in Refs.~\cite{Yamazaki:2011nd,Yamazaki:2012hi, Wagman:2017tmp,Yamazaki:2015asa,Beane:2011iw,Beane:2012vq,Beane:2013br,Orginos:2015aya,Berkowitz:2015eaa}.
Therefore conclusions in these references are not valid anymore, and 
more reliable methods such as the variational method\cite{Luscher:1990ck} are required to obtain the definite conclusion on the $NN$ system at heavy pion masses.

\section{Operator dependence}
As manifestations of  ``the fake plateau problem", 
there exist several symptoms,
one of which is the operator dependence of the plateau.
Fig.~\ref{fig:demo} (Right) shows $\Delta E_{NN}^{\rm eff}(t)$ obtained from wall and smeared sources, 
which indicates plateaux at different values\cite{Iritani:2016jie}, so that at least one of them must be fake.

Authors of Ref.~\cite{Yamazaki:2017euu} claimed that 
the plateau in $\Delta E_{NN}^{\rm eff}(t)$ from the wall source is 
fake, caused by 
a cancellation between the $NN$ effective energy and the nucleon effective mass, both of which seem to reach plateaux slower than $\Delta E_{NN}^{\rm eff}(t)$.    
This argument, however, does not properly address the issue in Ref.~\cite{Iritani:2016jie}  as explained below.

\begin{figure}[tb]
  \vskip -0.7cm
\centering
  \includegraphics[width=0.47\textwidth]{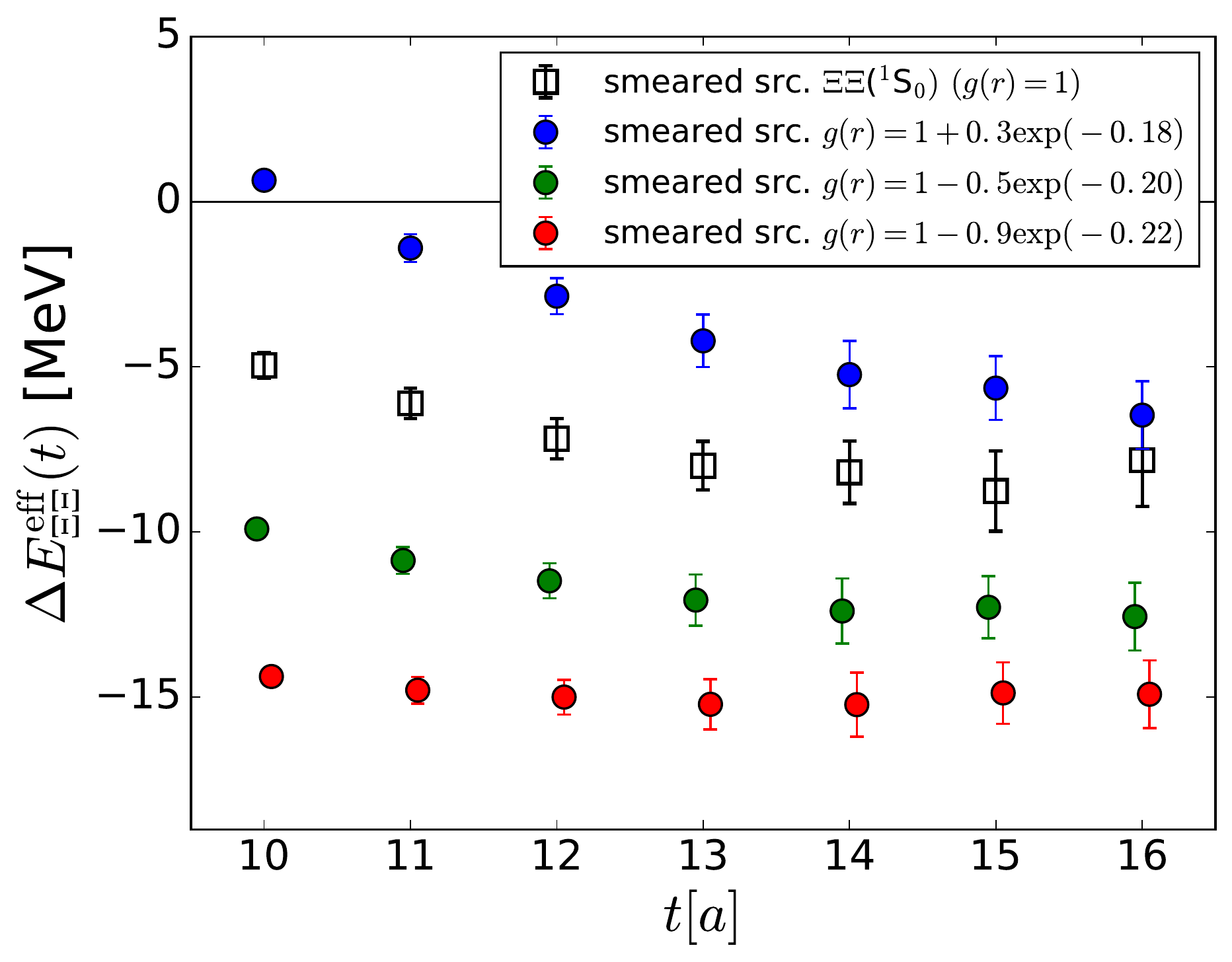}
  \includegraphics[width=0.47\textwidth]{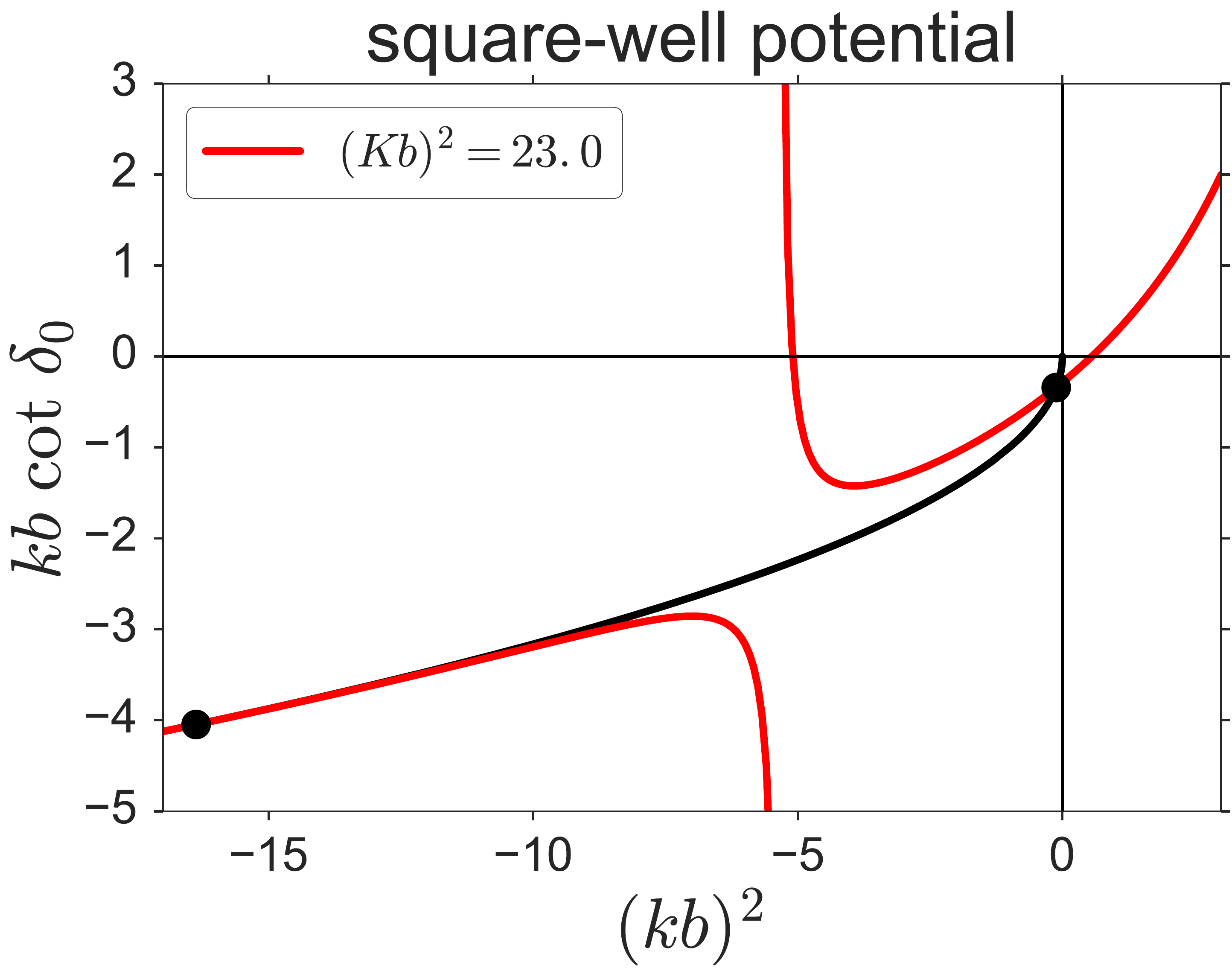}
 \caption{
(Left)  The effective energy shift $\Delta E^{\rm eff}_{\Xi\Xi}(t)$ from the smeared source with different sink operators\cite{Iritani:2016jie}. 
(Right)   An example of the physically allowed ERE compatible with two bound states\cite{Iritani:2016jie}.
}
 \label{fig:sink}
  \vskip -0.5cm
\end{figure}

First of all,  whether  the plateau from the wall source is correct or not does not logically
affect whether the plateau from the smeared source is correct or not.
Moreover 
there is no reason for the smeared source works better than the wall source for the $NN$ system,
since the smeared source is known to suppress  inelastic contributions to nucleon mass but 
to couple strongly to elastic scattering states with non-zero momenta. 
In fact, as shown in Fig.~\ref{fig:sink} (Left), where the effective energy of $\Xi\Xi (^1S_0)$ is plotted for several different sink operators,\footnote{$\Xi\Xi$ has smaller statistical errors than $NN$. See Appendix A in Ref.~\cite{Iritani:2016jie} for details.}
the smeared source produces several different plateaux, (almost) all of which should be fake.
This clearly shows that the plateau method does not have a predictive power at all.

In this conference\cite{Berkowitz},  Berkowitz and his collaborators pointed out 
a possibility that
inelastic excited state contributions in the nucleon effective mass may be largely cancelled by those in the $NN$ effective energy, so that the plateau may appear at  earlier $t$ in  $\Delta E_{NN}^{\rm eff}(t)$
as long as contaminations from elastic excited states in the $NN$ effective energy are negligible.
This possibility, however, should be verified by more reliable methods such as the variational method\cite{Luscher:1990ck}, since the simple plateau method cannot provide any information on this possibility.

In Ref.~\cite{Yamazaki:2017euu}, authors also claimed that the wall source and the smeared source agree at larger $t$ within large errors, as for data at $t\ge 16$ in the case of Fig.~\ref{fig:demo} (Left).
If so, however, one has to fit data at $t\ge 16$, leading to a different central value with much larger errors than before (solid black lines).  Moreover, even if the agreement between two sources is observed,
it does not guarantee that the results are reliable,
since $t$ is sill much smaller than the inverse of the typical excitation energy.

It is pointless to speculate a possible scenario for results from the plateau method to be correct
without clear evidences.
Instead one has to show that they are indeed correct without relying on the plateau method.
The issue is ``the fake plateau problem", not  the wall source against the smeared source unlike the argument in Ref.~\cite{Yamazaki:2017euu}. 

\section{Comment on Ref.~\cite{Beane:2017edf}}
\label{sec:reply}
Authors of Ref.~\cite{Beane:2017edf} claimed that the assessment in Ref.~\cite{Iritani:2017rlk} to NPL2013\cite{Beane:2012vq,Beane:2013br} is inadequate and NPL2013 has passed all tests in Tab.~\ref{tab:Summary}. In this section, we investigate their claims one by one and conclude that they are invalid or  irrelevant to address the fake plateau problem.

\subsection{Operator dependence}
Authors of Ref.~\cite{Beane:2017edf} compared the ground state energy of the $NN$ system in $^1S_0$ and $^3S_1$ channels among NPL2013\cite{Beane:2012vq,Beane:2013br}, CalLat2017\cite{Berkowitz:2015eaa} and NPL2017\cite{Tiburzi:2017iux}, all of which employed the same gauge ensembles, and concluded that the source operator dependence  is not established,
since the $\chi^2$ test on the assumption that all results are same gives $P$-values (significances) ranging from 8\% to 65\%. 

However, comparing all data at once  is not an optimal way to investigate the operator dependence.
Instead, it is better to perform the one-to-one comparison for a maximal detection of the dependence.
Such comparisons are made for NPL2013 and CalLat2017 in Tab.~\ref{tab:1to1},
where the operator dependence of the ground state energy is clearly seen between the center of mass and the moving systems in $^1S_0$ of NPL2013, between non-displaced and displaced source operators of CalLat2017, and
between NPL2013 and CalLat2017 (displaced), while that of the 1st excited state is confirmed between NPL2013 and CalLat2017  for $^3S_1$ at $L=32$.
\begin{table}[tbh]
\centering
\caption{$\chi^2/$dof ($P$-value \%) between two data  using NPL2013 configurations.
(Top) The ground state energy from
NPL(0,2) and CL(0,d), where NPL(0) and NPL(2) means NPL2013 \cite{Beane:2013br,Beane:2012vq} in the center of mass and the moving systems, respectively,  
while CL(0) and CL(d) are CalLat2017\cite{Berkowitz:2015eaa}  with non-displaced and displaced source operators, respectively.
The upper-right (lower-left) triangle corresponds to $L=24(32)$. 
As possible correlations between two data are neglected here, $\chi^2$ in this table might be underestimated.
(Bottom) The 1st excited state energy from NPL2013(NPL) and CalLat2017(CL).
 }  
\label{tab:1to1}
\begin{tabular}{|c||c|c|c||c|c|c|c|} 
\hline
& \multicolumn{3}{c||}{$NN(^1S_0)$} & \multicolumn{4}{c|}{$NN(^3S_1)$} \\ 
\hline
 & NPL(0) &NPL(2) &CL(0) & NPL(0) &NPL(2) &CL(0) &CL(d) \\
\hline
NPL(0) & - &    3.7 (5\%)    &   0.3(56\%)    &  - &  2.4(12\%)        &    0.4(52\%)     & 0.5(47\%)    \\  
NPL(2)  & 4.2(4\%) & - &   2.6(11\%)        & 2.7(10\%)&  - &   1.1(31\%)     & 5.3 (2\%)    \\
CL(0)  & 0.3(59\%) & 2.5(11\%)  &   -         &   1.5(22\%)& 0.5(50\%) & -  &     2.4(12\%)     \\
CL(d)   & 5.0 (3\%)   &  17(<1\%)  & 8.6(<1\%)      &  11(<1\%)  &   24(<1\%)    &    31 (<1\%)         &  -    \\
\hline
\end{tabular}
\begin{tabular}{|c||c|c||c|c|c|} 
\hline
 & $NN(^1S_0)$ $L=24$ & $NN(^1S_0)$ $L=32$ & $NN(^3S_1)$ $L=24$ & $NN(^3S_1)$ $L=32$ \\
 \hline
 NPL-CL &  0.04(84\%) & 2.6(11\%) & 0.44 (51\%) & 8.0 (0.5\%) \\
 \hline
\end{tabular}
\vskip -0.3cm
\end{table}
 
Furthermore, an absence of the operator dependence for the limited sets of operators, even if it were confirmed, does not mean the absence of ``the fake plateau problem". One has to prove separately that the plateau is not fake.

It is also claimed in Refs.~\cite{Berkowitz:2015eaa,Beane:2017edf}, without any numerical evidences,
that the result from the displaced source operator in CalLat2017 does not corresponds to the ground state energy, so that no operator dependence exists in CalLat2017.\footnote{
The argument in Ref.~\cite{Berkowitz:2015eaa} that the displaced operator does not couple to the ground state does not have solid ground, since it assumes the absence of the fake plateau problem without evidence. }
This interpretation, however, brings other problems.
If the plateau of the displaced operator correspond to the excited scattering state,
the negative energy shift of this state disagrees with the positive value from other sources with non-zero momentum. Thus it creates another huge operator dependence.
If this plateau corresponds to the second shallow bound state, on the other hand, 
one has to explain why NPL2013 missed this state, and 
the effective range expansion (ERE) in NPL2013 and CalLat2017 should be drastically modified as in Fig.~\ref{fig:sink} (Right), to satisfy the physical pole condition\cite{Iritani:2017rlk}.

As we confirmed above, NPL2013/CalLat2017 data show the source dependence,
which strongly suggests that 
their extractions of energies  from plateaux suffer from ``the fake plateau problem".

Refs.~\cite{Beane:2017edf}  also claimed that data at the largest volume ($L=48$) is important to determine the binding energy in the infinite volume extrapolation.
Although this statement itself is correct, the large volume data have nothing to do with the validity of data at smaller volumes discussed in Ref.~\cite{Iritani:2017rlk}.

\subsection{Volume scaling of energies}
Showing that $NN$ energies in NPL2013 are almost volume independent, 
authors of Ref.~\cite{Beane:2017edf} claimed that this volume independence rules out the possibility that the plateau structures are caused by the cancelations among excited states.
Contrary to this claim, however, in general there is no logical connection between the volume independence and the correctness of the plateau.
Indeed, while energies in YKU2012\cite{Yamazaki:2012hi} are almost volume independent, the corresponding ERE is very singular as shown in Ref.~\cite{Iritani:2017rlk} and the source/sink operator dependences are observed as seen in the previous section.
This means that the volume independence does not guarantee the correctness of data, contrary to the claim of Ref.~\cite{Beane:2017edf}.

\subsection{Physical pole condition}
Authors of Ref.~\cite{Beane:2017edf} argued that ERE fits in NPL2013 satisfy the physical pole condition 
\begin{eqnarray}
\left.
\frac{d}{d k^2}\left[k\cot\delta_0(k) - (-\sqrt{-k^2})\right]\right.\vert_{k^2= -k^2_b} < 0,
\label{eq:phy_pole}
\end{eqnarray}
where $k^2_b$ correspond to the  bound state pole,  by showing that the slope of $ -\sqrt{-k^2}$
at $k^2=-k^2_b$ is larger than the slope of ERE, called the effective range.
\begin{figure}[tb]
 \vskip -0.5cm
\centering
  \includegraphics[width=0.47\textwidth]{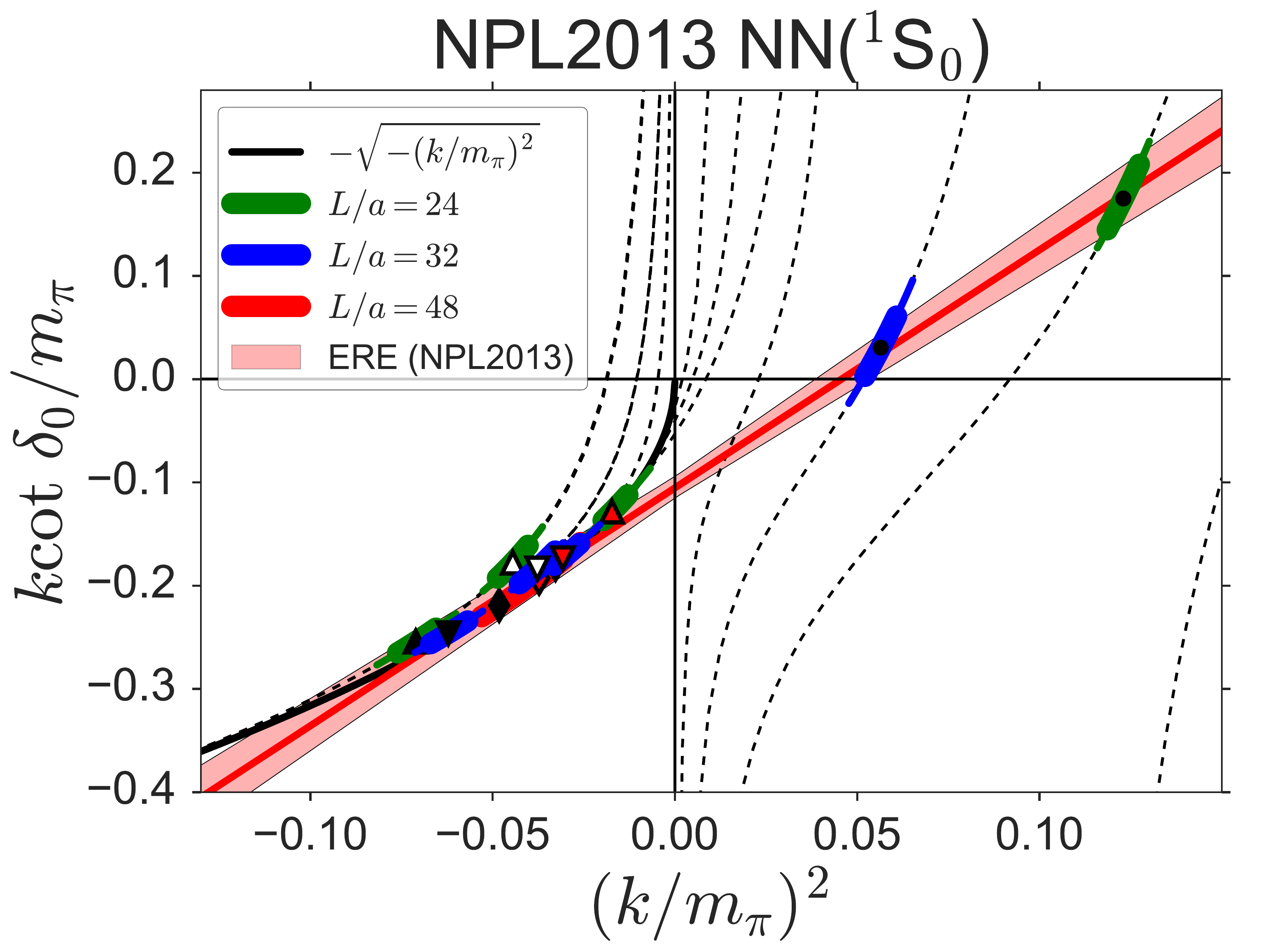}
  \includegraphics[width=0.47\textwidth]{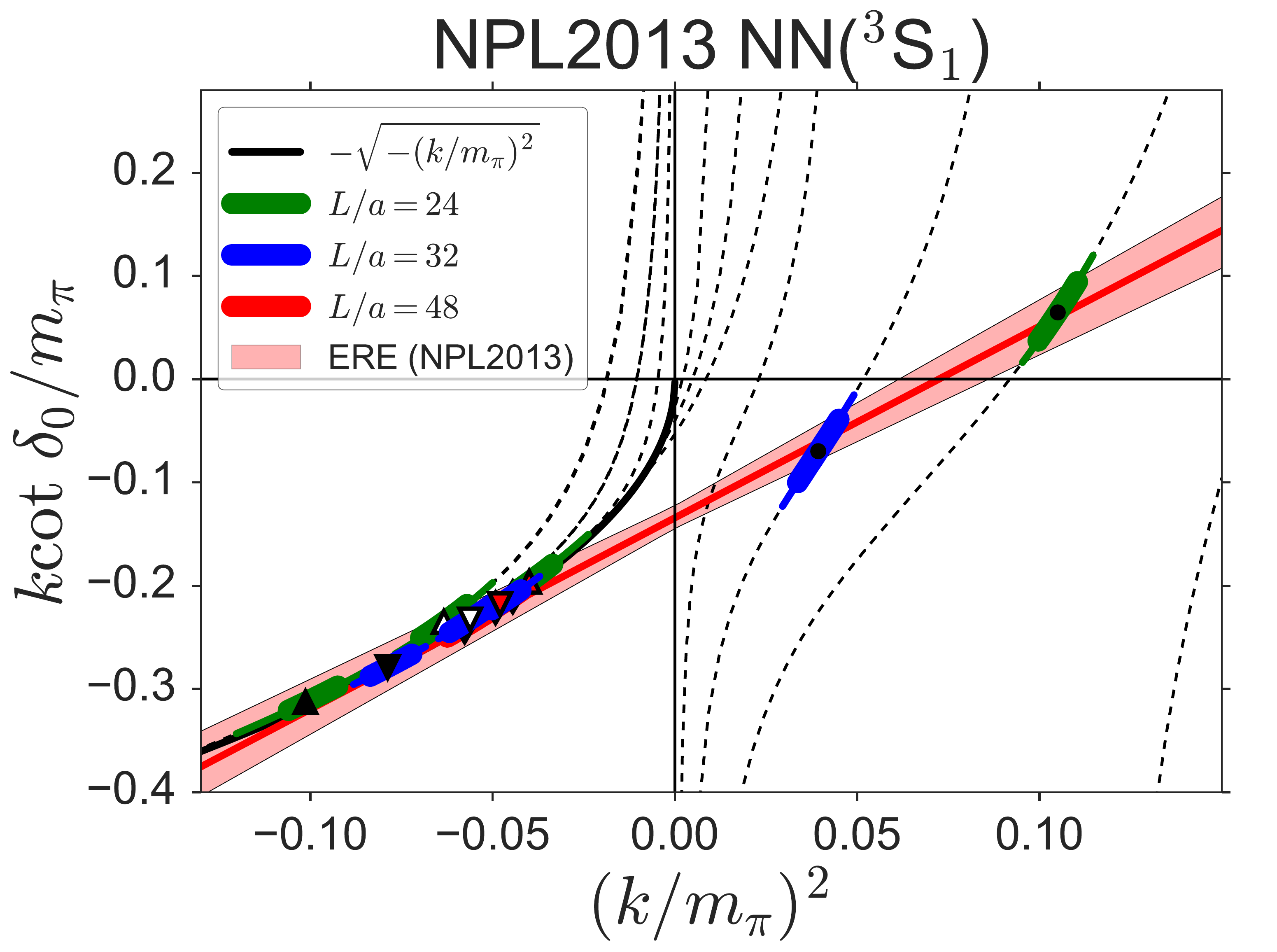}
 \caption{
$k\cot\delta_0(k)/m_\pi$ as a function of $(k/m_\pi)^2$ for $NN (^1S_0)$ (Left) and $NN (^3S_1)$
for NPL2013 data. 
The black solid line represents the bound state condition $-\sqrt{-(k/m_\pi)^2}$, while  NLO ERE fits are shown by red lines, together with light red error bands\cite{Iritani:2017rlk}. 
}
 \label{fig:ERE}
  \vskip -0.6cm
\end{figure}  

This argument, however, is insufficient to check the physical pole condition,
which implies that {\it all} intersection points between the ERE and the pole condition satisfy eq.~(\ref{eq:phy_pole}).
Fig.~\ref{fig:ERE} shows $k\cot\delta_0(k)/m_\pi$ as a function of $(k/m_\pi)^2$, together with the NLO ERE line (as well as its error band),  taken directly from figures in NPL2013 and replotted in Ref.~\cite{Iritani:2017rlk}.
As the red line in both channels appears below the black solid line at $(k/m_\pi)^2 \simeq 0$ and
$(k/m_\pi)^2 \simeq -0.13$, the ERE line must intersect with the pole condition twice, 
so that the deeper one inevitably violates the physical pole condition.  We therefore put `No' on sanity check (iii) for $NN(^1S_0)$ and $NN(^3S_1)$ in the original assessment~\cite{Iritani:2017rlk}.
\footnote{Since the light red error band goes above the pole condition  at $(k/m_\pi)^2 \simeq -0.13$ for $NN(^3S_1)$, we will change `No' to `?' in the revised version of Ref.~\cite{Iritani:2017rlk}.}

Ref.~\cite{Beane:2017edf} states that ``The incorrect conclusions reached in HAL potentially arise from attempting to reanalyze highly correlated data, such as the encountered in the phase-shift analysis, without making use of those correlations".\footnote{
After the lattice conference, the version 2 of \cite{Beane:2017edf} appeared without
this sentence.}
However, this statement is not correct, since the ERE lines in the figures are NOT our reanalysis but are taken from the figure in  NPL2013, as already mentioned. 

\section{Some other concerns on Ref.~\cite{Beane:2017edf}}

\subsection{Inadequate data handlling}
In Ref.~\cite{Beane:2017edf}, authors stated that they made new ERE analyses to their data
and claimed that data passed the sanity check.
As already emphasized, our sanity check in Ref.~\cite{Iritani:2017rlk} was applied to ERE fits in NPL2013 but not to new analysis, so that the statement in Ref.~\cite{Beane:2017edf} that
`` the results of Refs.[6.7]\footnote{Refs.[6,7] correspond to NPL2013.} pass this check, in stark contrast to the claims of HAL" 
is inadequate. 
Moreover, we found that  values of the energy shift were modified from NPL2013,
as summarized in Tab.~\ref{tab:comp}, where data in NPL2013 are compared with data read off from figures in Ref.~\cite{Beane:2017edf}. At this conference,
we pointed out  that criticisms based on data modified from the original ones
cannot be taken as face value.\footnote{
Although this data modification is mentioned in its version 2, there still remains a similar statement 
that ``the results of Refs.[12.13] pass this check, contrary to the claims in HAL", where Refs.[12.13] mean NPL2013.} 
\begin{table}[tbh]
\centering
\caption{$E_{NN}-2 m_N$ (MeV) for the ground state (upper) and the first excited state (lower)
for various volumes in NPL2013 and Ref.~\cite{Beane:2017edf}, where $d$ denotes the total momentum $P =2\pi d /L$.
 }  
\label{tab:comp}
\begin{tabular}{|c|c||c|c|c||c|c|c|} 
\hline
&& \multicolumn{3}{c||}{$NN(^1S_0)$} & \multicolumn{3}{c|}{$NN(^3S_1)$} \\ 
\hline
$d$ & L & 24 & 32 & 48 & 24 &32 & 48 \\
\hline
0 & NPL2013 & -17.8(3.3)  &-15.1(2.8) &-13.1(5.2) & -25.4(5.4) &-22.5(3.5) &-19.7(5.2) \\ 
0 & \cite{Beane:2017edf} & -19.0($^{2.5}_{2.2}$) & -18.5($^{2.4}_{2.1}$)   &   -19.9($^{3.4}_{4.3}$)    &  -28.5($^{1.7}_{1.7}$) &  -26.3($^{2.7}_{2.2}$)&-25.3($^{3.9}_{4.9}$)       \\
2 &  NPL2013 & -28.5(4.5) & -24.9(3.8) &-19.3(4.4) & -40.7(8.3) & -31.6(4.2) &-23.1(6.8) \\
2 &  \cite{Beane:2017edf} & -23.5($^{3.3}_{2.7}$) &  -21.4($^{2.3}_{2.4}$) & -21.8($^{4.2}_{5.5}$) &
 -34.3(3.6) &  -29.6($^{3.3}_{2.8}$) &  -27.5($^{4.4}_{5.4}$) \\
 \hline\hline
0 & NPL2013 & 48.7(2.8)  &22.5(3.5) & N/A & 41.6(3.8) &15.7(3.8) & N/A \\ 
0 & \cite{Beane:2017edf} & 49.3($^{3.0}_{2.4}$) & 21.3(2.4)   & N/A  &  37.5($^{3.8}_{3.7}$) &  12.5($^{3.2}_{2.9}$)&   N/A    \\
2 &  \cite{Beane:2017edf} & 44.0($^{4.3}_{3.4}$) &  18.7($^{2.4}_{2.7}$) &  N/A &
 28.9($^{5.2}_{4.5}$) &  8.4($^{2.4}_{2.1}$) &  N/A \\ 
\hline
\end{tabular}
\end{table}

\subsection{Incorrect fitting}
  \vskip -0.3cm
\begin{figure}[tb]
\centering
  \includegraphics[width=0.47\textwidth]{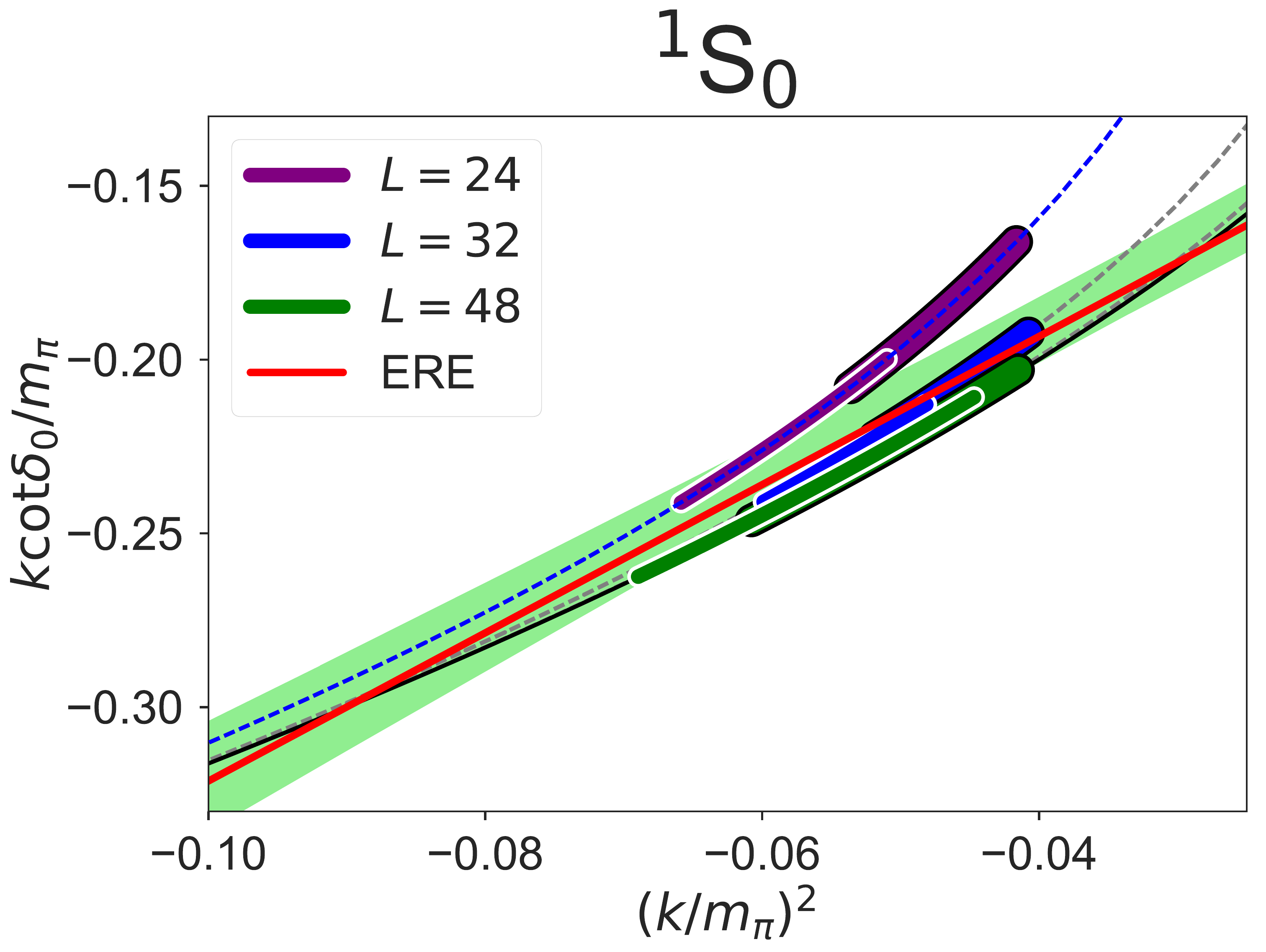}
  \includegraphics[width=0.47\textwidth]{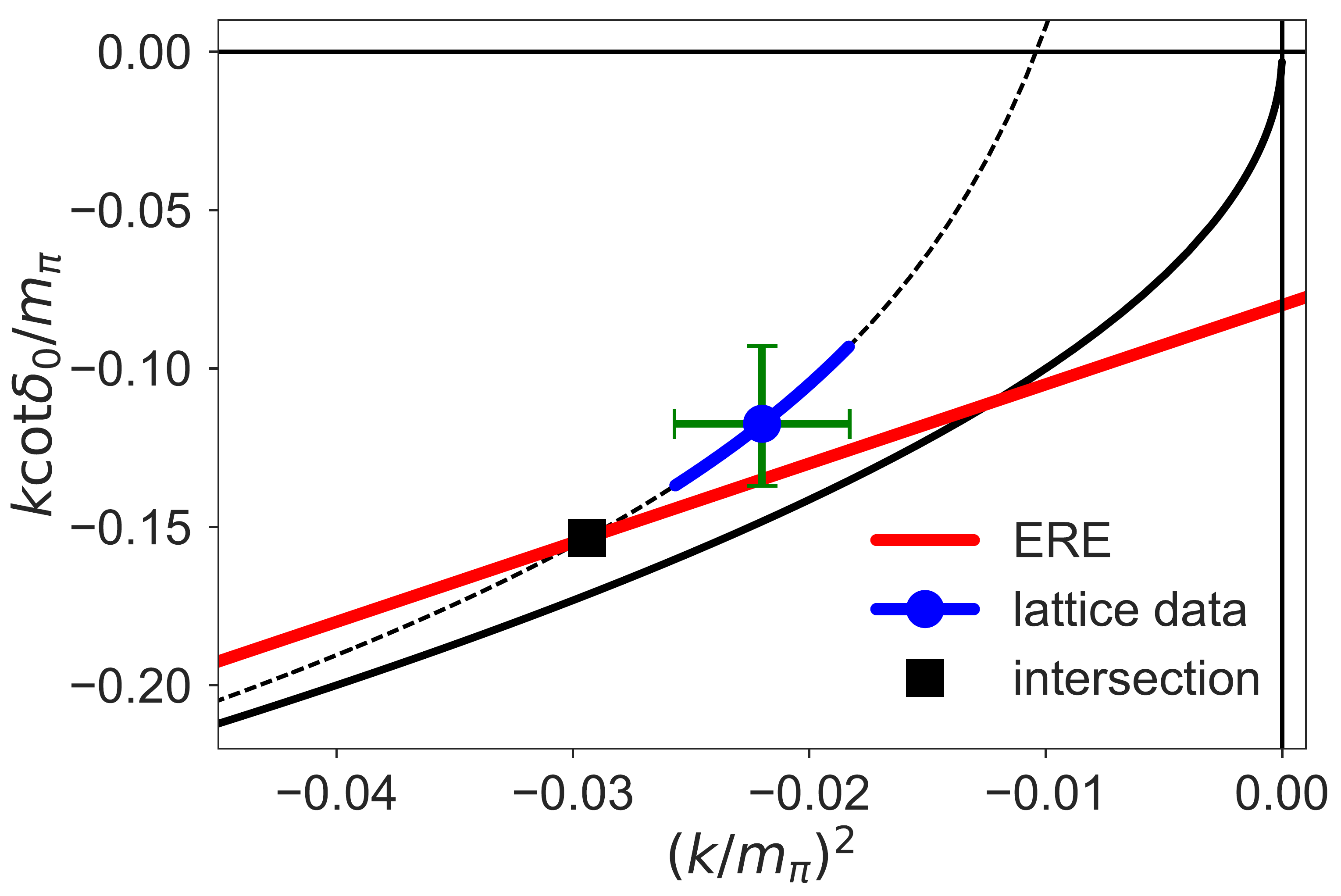}
 \caption{
(Left)  The ERE (red line) for $^1S_0$ of  Refs.~\cite{Beane:2017edf,Wagman:2017tmp} and
the finite volume formula (dashed lines).
(Right) The correct error analysis with the constraint by the L\"uscher's formula.
}
 \label{fig:ERE_fit}
  \vskip -0.5cm
\end{figure}  
The ERE for $^1S_0$ in Refs.~\cite{Beane:2017edf,Wagman:2017tmp} does not intersect with the line of the L\"uscher formula at $L=24$, as shown in Fig.~\ref{fig:ERE_fit} (Left).  Since the $\chi^2$ cannot be defined in this case, it is not clear how authors of Ref.~\cite{Beane:2017edf} obtained the line and its errors in their ERE analysis.
Since the vertical error is derived from the horizontal one,
one needs to fit lattice data by the ERE with the constraint due to the L\"uscher's finite volume formula,
as demonstrated in Fig.~\ref{fig:ERE_fit} (Right):
Once the ERE line is fixed by given fit parameters,  an intersection between the ERE and the L\"uscher's formula on each $L$ is determined (black square), and then $(k/m_\pi)^2$ of the intersection is compared with that of lattice data (blue circle) to calculate the corresponding $\chi^2$. 
As the correct $\chi^2$ cannot be defined without intersection,
it seems that Ref.~\cite{Beane:2017edf,Wagman:2017tmp} did not take into account this constraint correctly.

\subsection{Lattice spacing and pion mass}
NPLQCD collaboration employed the SU(3) flavor symmetric QCD configurations for
various publications\cite{Beane:2012vq, Beane:2013br,Beane:2013kca,Beane:2014ora,Beane:2014sda,Beane:2015yha,Chang:2015qxa,Detmold:2015daa,Parreno:2016fwu,Savage:2016kon,Shanahan:2017bgi,Wagman:2017tmp,Tiburzi:2017iux,Beane:2017edf}.
While the lattice spacing $a$ in Refs.~\cite{Beane:2012vq,Beane:2013br,Beane:2013kca,Beane:2014sda,Parreno:2016fwu,Beane:2017edf,Wagman:2017tmp} ($a\simeq 0.15$ fm)
is different from $a$ in Refs.~\cite{Beane:2014ora,Beane:2015yha,Chang:2015qxa,Detmold:2015daa,Savage:2016kon,Shanahan:2017bgi,Tiburzi:2017iux} ($a\simeq 0.12$ fm), 
the pion mass $m_\pi$ in these references is always the same ($m_\pi\sim 806$ MeV).
Although the lattice spacing can depend on the physical input ($r_0$, light hadron masses,
heavy meson spectra, etc.), the pion mass (and dimensionfull quantities such as the energy shift) must be  changed accordingly. 
It is not clear as well how these values are obtained and how the chiral and/or continuum extrapolation are made.

\section{Summary}
As we discussed in this report, our sanity check in Tab.~\ref{tab:Summary} remains valid, contrary to the claims in Refs.~\cite{Yamazaki:2017euu,Beane:2017edf}, which contain either invalid statements or inadequate data handling or both.

\bibliography{lattice2017_47_aoki}

\end{document}